\documentclass[
reprint,
bibnotes,
amsmath,amssymb,
aps,
prb,
]{revtex4-1}

\usepackage{graphicx}
\usepackage{dcolumn}
\usepackage{bm}
\usepackage{color}
\usepackage{hyperref}
\usepackage{cleveref}
\usepackage{caption}
\usepackage{subfig}
\usepackage{float}

\begin{document}
\title{Surface Melting and Break up of Metal Nanowires: Theory and Molecular Dynamics Simulation}

\author{Kannan M. Ridings}
\email{k.ridings@auckland.ac.nz}
\affiliation{%
MacDiarmid Institute for Advanced Materials and Nanotechnology, Department of Physics, University of Auckland, New Zealand\\
}
\author{Thomas S. Aldershof}
\email{t.aldershof@uq.edu.au}
\affiliation{%
School of Mathematics and Physics, University of Queensland, Australia\\
}

\author{Shaun C. Hendy}
\email{shaun.hendy@auckland.ac.nz}
\affiliation{%
Te P\={u}naha Matatini, Department of Physics, University of Auckland, New Zealand\\
}

\date{\today}

\begin{abstract}
We consider the surface melting of metal nanowires by solving a phenomenological two-parabola Landau model and by conducting molecular dynamics simulations of nickel and aluminium nanowires. The model suggests that surface melting will precede bulk melting when the spreading parameter $\Delta \gamma$ for the melt in contact with the solid surface is positive (i.e. if the melt wets or partially wets the surface) and the wire is sufficiently thick, as is the case for planar surfaces and sufficiently large nanoparticles. Surface melting does not occur if $\Delta \gamma$ is negative. We test this model, which assumes the surface energies of the wire are isotropic, using molecular dynamics simulations. For nickel, we observe the onset of anisotropic surface melting associated with each of the two surface facets present, but this gives way to uniform surface melting and the solid melts radially until the solid core eventually breaks up. For aluminium, while we observe complete surface melting of one facet, the lowest energy surface remains partially dry even up to the point where the melt completely penetrates the solid core.

%
\end{abstract}

\maketitle
\section{Introduction}\label{Intro}
Nanostructured materials typically have lower melting points than that of the bulk due to their high surface area to volume ratios, which reduces their stability relative to the molten phase \cite{wronski1967size, coombes1972melting, di1995maximum}. Indeed, with the exception of very small clusters \cite{breaux2003hot, steenbergen2012electronic} or particles with non-melting surfaces \cite{hendy2009superheating}, the melting point depression in metal nanoparticles is found to scale in proportion to this ratio \cite{bachels2000melting}. The melting of nanowires has been less studied \cite{samanta2014microscopic, wu2015self}, despite their technological relevance, in part because wires are not thermodynamically stable. Plateau and Rayleigh in the late 19th century showed that a liquid cylinder of radius $r$ will become unstable to radial perturbations of wavelengths $\lambda$ which exceed the circumference of the cylinder, driven by the corresponding reduction surface energy. A similar phenomenon can occur in metal nanowires, where the ratio of surface area to volume is very large. Indeed, it has been shown in experiment that the fragmentation of nanowires into a chain of spheres can occur at elevated temperatures\cite{dutta2014silico} via a Rayleigh-type  instability\cite{toimil2004fragmentation,shin2007size} as the wire seeks to reduce its surface area.

In metal wires, surface melting may also play a role in this break up process \cite{wu2015self}. In bulk metals, melting typically initiates at a surface at temperatures below that of the bulk melting temperature \cite{Observation_of_surface_melting, pluis1990surface}. Metal surfaces that satisfy $\Delta \gamma = \gamma_{sv} - \gamma_{lv} - \gamma_{sl} > 0$ (where $\gamma_{sv}$, $\gamma_{lv}$, $\gamma_{sl}$ are the interfacial energies of the solid-vapour (sv), liquid-vapour (lv) and solid-liquid (sl) interfaces) can reduce their surface energy by melting at a temperature below the bulk melting temperature. $\Delta \gamma$ is sometimes called a spreading parameter: for surfaces with $\Delta \gamma > 0$ then the melt will wet or partially wet the corresponding solid surface. Surface melting is also know to occur in nanoparticles and nanowires if they are bounded by surfaces with $\Delta \gamma > 0$. Once surface melting is initiated on a wire, it may act to accelerate the break up process. Indeed, in a recent molecular dynamics study, the solid core of an aluminium wire was observed the break up during the melting of the wire \cite{wu2015self}. 

Nonetheless, not all metal surfaces pre-melt \cite{carnevali1987melting}. Surfaces with $\Delta \gamma < 0$, such as the Al(111), Pb(111) and Al(100) planes\cite{van1990melting,NM_surfs,pluis1987crystal} can remain solid up to the bulk melting temperature. Such surfaces are called non-melting (NM), and one would expect nanowires bounded by NM surfaces to melt and break up very differently to those bounded by melting surfaces. They may well be more stable both to melting and to break up than wires with surfaces that pre-melt. Despite the likely importance of surface melting phenomena, models of nanowire melting typically neglect surface melting phenomena, even though they often take into account the reduced melting temperature of wires relative to the bulk (see for instance \cite{wu2015self,Florio2016}). 

The purpose of this paper is to investigate both the surface melting of metal nanowires and its potential influence on nanowire break up. In particular, we will examine the surface melting transition in nickel and aluminium nanowires, bounded by both melting and non-melting surfaces, using a phenomenological Landau-type model and molecular dynamics simulations. The double-parabola Landau-type model has previously been applied to planar surfaces \cite{pluis1990surface} and to spherical particles \cite{sakai1996surface, Chang2005surface}, but we believe it is the first application of such a model to a wire. In the first section we show that this model, which assumes the surface energies of the wire are isotropic, predicts that surface melting will precede bulk melting when $\Delta \gamma$ is positive (i.e. if the melt wets or partially wets the surface), as is the case for planar surfaces and sufficiently large nanoparticles. In subsequent sections, we test this model using molecular dynamics simulations of nickel wires, bounded by melting surfaces, and then aluminium nanowires, bounded by both melting and non-melting surfaces. Finally, using the molecular simulations, we investigate the break-up of the solid in the presence of surface melting for both nickel and aluminium.

\section{Phenomenological model for surface melting}\label{S: F_eval}
In this section we solve a phenomenological Landau-type model to describe melting point depression and surface melting in a metal nanowire. Pluis et al \cite{pluis1990surface} developed such a model to describe surface melting at planar metal surfaces. This has been extended to describe surface melting in spherical particles \cite{sakai1996surface, Chang2005surface}, but to the best of our knowledge, this approach has not been applied to study surface melting in nanowires. In the limit of small surface curvature, one can appeal to the model of Pluis et al \cite{pluis1990surface} to describe surface melting in a cylindrical geometry, but as the radius of curvature approaches the width of the solid-liquid interface the value of such an approximation is less clear. For this reason, it is useful to solve the double-parabola Landau model for surface melting in a cylindrical geometry. 

We start by defining a Landau free energy functional $F$ (per unit length) for a cylindrically symmetric wire. Assuming isotropic material properties (which in general for nanowires is not true but otherwise simplifies the model) we express the solid state of a nanowire of radius $R$ in terms of a crystalline order parameter $M(r)$ \cite{pluis1990surface}:
\begin{multline}
F \left[ M(r) \right] = \\ 2 \pi \int_{0}^{R} r \left\lbrace f\left( M \right) + \frac{J}{2} \left( \frac{dM}{dr} \right)^2\right\rbrace dr + 2 \pi R f_s\left( M \right) \label{FEeqn}
\end{multline}
where $f$ and $f_s$ are the bulk and surface free energies per unit volume respectively, while $J = 4 \gamma_{sl} \xi$ is a parameter (taken here to be independent of temperature) that is proportional to the correlation length $\xi$ at the solid-liquid interface $\gamma_{sl}$. By taking the correlation length $\xi \rightarrow 0$ in equation~\ref{FEeqn} the free energy $F\left[ M(r) \right]$ recovered will simply be the classical result.

\begin{figure}[htp]
\resizebox{\columnwidth}{!}{\includegraphics[width=0.45\textwidth]{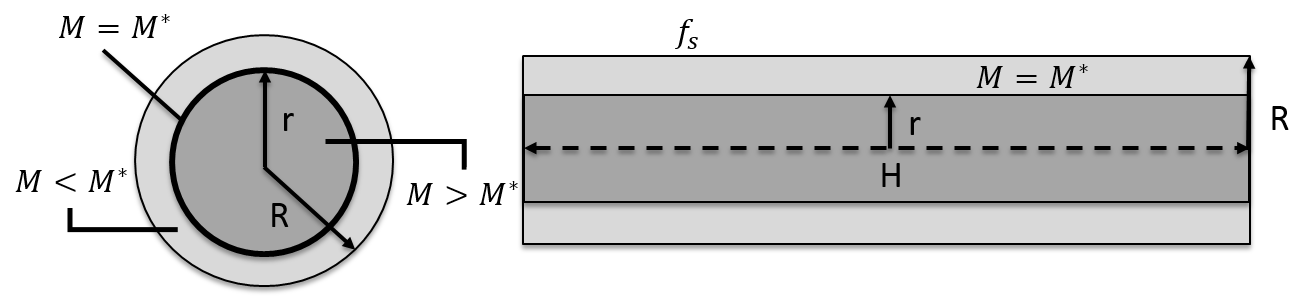}}
\caption{\label{Fig:Surf_melt} A surface melted nanowire of radius $R$ and length $H$. The thickness of the melt is $R-r$ with $r$ being the solid core radii. The solid and liquid phases are in the vicinity of the wire where $M>M^{*}$ and $M<M^{*}$ respectively with the phase boundary $M=M^{*}$ indicated.}
\end{figure}

The bulk free energy $f$ at temperature $T$ is approximated piecewise with a pair of parabolas:
\begin{equation}
f(M) = \begin{cases} 
\frac{\alpha}{2} M^2 + \Lambda(T) & \mbox{for } M < M^* \\
\frac{\alpha}{2} \left(1-M\right)^2 & \mbox{for } M > M^* \end{cases} 	
\end{equation}
The material dependent parameter $\alpha = 4 \gamma_{sl}/\xi$ is also related to $\gamma_{sl}$ and the correlation length $\xi$. $\Lambda (T) \simeq \rho L \left(\frac{T_c-T}{T_c} \right)$ where $T_c$ is the bulk melting temperature and $\rho L$ is the bulk latent heat of melting (by volume). An intersection at $M^*$ (where $M^* = \frac{1}{2} - \frac{\Lambda(T)}{\alpha}$), defines the boundary between the solid $(M > M^*)$ and liquid phases $(M^* < M)$ as shown in Fig.~\ref{Fig:Surf_melt}. The surface free energy is also modelled as a parabola:
\begin{equation}
f_s = \frac{\alpha_s}{2} M\left(R\right)^2 + \gamma_{lv} 
\end{equation}
where $\alpha_s$ is given by $\alpha_s = 4 \gamma_{sl} \left(\frac{1+\Delta \gamma/\gamma_{sl}}{1-\Delta \gamma/\gamma_{sl}}\right).$

At equilibrium, $\delta F / \delta M = 0$, which requires that $M$ satisfy the following differential equation
\begin{equation}
\frac{d^2 M}{dr^2} + \frac{1}{r} \frac{dM}{dr} = \frac{1}{\xi^2} \begin{cases} M & \mbox{for } M < M^* \\
M-1 & \mbox{for } M > M^* \end{cases} 	
\label{MBesselEqn}
\end{equation}
subject to a boundary condition at the surface of the wire $r=R$:
\begin{equation}
\left. J \frac{dM}{dr} \right|_{r=R} = - \left. \frac{\partial f_s}{\partial M} \right|_{r=R}.
\end{equation}

When $M(R) \geq M^*$, which is the case prior to the onset of surface melting, the solution of (\ref{MBesselEqn}) is given by:
\begin{equation}
M(r) = 1 - \frac{I_0 \left( r/\xi \right)}{I_0 \left( R/\xi \right) + \kappa I_1 \left( R/\xi \right)}
\label{OrderSolid}
\end{equation}
where $I_0$ and $I_1$ are modified Bessel functions of the first kind and $\kappa = J/\xi \alpha_s$. The melting temperature of the wire $T_m$ is that where the free energy of the liquid wire is equal to that of the solid wire ($F \left[ 0 \right] = F \left[M \right]$ for $M(R)>M^*$) which yields
\begin{equation}
\label{eqn:Tmelt}
T_m = T_c\left(1 - \frac{4 \gamma_{sl}}{\rho L R} \left(\frac{I_1 \left( R/\xi \right)}{I_0 \left( R/\xi \right) + \kappa I_1 \left( R/\xi \right)} \right)\right)
\end{equation}
Note that the solution The onset of surface melting occurs at a temperature $T_s$ where the order parameter at the wire surface $M(R)$ becomes equal to $M^*$. Solving $M(R)=M^*$ gives the temperature $T_s$ at which surface melting begins:
\begin{equation}
\label{eqn:Tsurf}
T_s = T_c\left(1 + \frac{4 \gamma_{sl}}{\rho L \xi} \left( \frac{1}{2} -\frac{I_0 \left( R/\xi \right)}{I_0 \left( R/\xi \right) + \kappa I_1 \left( R/\xi \right)} \right)\right)
\end{equation}
When $\xi \ll R$, we obtain
\begin{equation}
\label{Eq:Tmelt_O2}
T_m = T_c\left(1 - \frac{2(\gamma_{sv}-\gamma_{lv})}{\rho L R}\left(1 + \frac{\gamma_{sv}-\gamma_{lv}}{4 \gamma_{sl}}\frac{\xi}{R} + ... \right) \right)
\end{equation}
and
\begin{equation}
\label{Eq:Tsurf_O2}
T_s = T_c\left(1 - \frac{\gamma_{sl}}{\rho L \xi}\left(\frac{2 \Delta \gamma}{\gamma_{sl}} + \frac{\xi}{2 R}\left(1 - \left(\Delta \gamma/\gamma_{sl}\right)^2 \right) + ... \right)\right)
\end{equation}
to first order in $\xi/R$. Note that in the limit of large curvature ($\xi/R \rightarrow 0$), $T_s = T_c (1-2 \Delta \gamma/\rho L \xi)$, which coincides with the surface melting temperature computed by Pluis et al \cite{pluis1990surface} for a planar surface. We conclude that provided $R \gg \xi$, the melting point and surface melting temperature of a nanowire is inversely proportional to the radius of the wire as would be expected. 

Finally, we see that surface melting will only occur when $T_s < T_m$. This is not the case when $\Delta \gamma <0$, where it can be shown that $T_s > T_m$, so the onset of complete melting will occur before any surface melting. However, surface melting will occur prior to complete melting if $\Delta \gamma >0$ and $R > R_c$, where $R_c$ is a critical radius given by the implicit equation:
\begin{equation}
\frac{R_c}{2 \xi}\left(\frac{I_0\left(R_c/\xi\right)}{I_1\left(R_c/\xi\right)} - \kappa \right) =1
\label{Eq:Rc_Fits}
\end{equation}
Wires with a radius less than $R_c$ will melt  completely prior surface melting, which is similar to what is predicted by the double-parabola model when applied to small spherical particles (e.g. see \cite{Chang2005surface}).
%
%

\section{Computational Details}\label{S:Comp_Details}
In this section we detail the appraoch taken in our molecular dynamics simulations. Periodic boundary conditions in the direction the wire was oriented allowed us to simulate an infinitely long wire, as well suppressing long-wavelength instabilities that otherwise might cause the wire to break prior to complete melting. The wires were chosen to be bounded by $\{100\}$ and $\{110\}$ surfaces and the atomistic structure was constructed using a Wulff-type construction of the form $\sum_j (L_j - \lambda \gamma_j)A_j = 0$ where $\lambda = L_j/\gamma_j$ (where the $\gamma_j$ were obtained from table~\ref{tab:critical_R}). The distance from the centre of the wire to the $\{110\}$ surface, $L_{\{110\}}$ is then proportional to the ratio of the surface energies of the two exposed crystal faces: $L_{\{110\}} = L_{\{100\}}(\gamma_{\{110\}}/\gamma_{\{100\}})$. When we refer to a wire radius $R$ in what follows, we will take this to mean $L_{100}$.

\begin{table}[htp]
\begin{tabular*}{\columnwidth}{@{\extracolsep{\stretch{1}}}*{7}{r}@{}}
\hline
    		Metal		& $\gamma_{sv}$ & $\gamma_{lv}$	&$\gamma_{sl}$ & $\Delta \gamma$	& $\xi$( \AA\ ) \\
		\hline
        Nickel			 &0.1425\cite{tyson1977surface}	& 0.1149\cite{gammas_lv} 	& 0.0217\cite{gammas_sl_exp} & 0.00590	& 10.5\cite{pluis1990surface}	\\
        Aluminium		 	 &0.0613\cite{murr1975interfacial} 	& 0.0542\cite{gammas_lv}	& 0.00750\cite{gammas_sl_exp} & -0.000630	& 9.96\cite{pluis1990surface}\\
        \hline        
 	\end{tabular*}  
\begin{flushleft}
\captionsetup{justification=raggedright}
\caption{\label{tab:critical_R}
 Interfacial free energy densities given in $eV/$\AA$^{2}$ and $\xi$ is given angstroms \AA. Values for $\gamma_{sv}$ are calculated via a semi-theoretical approach, $\gamma_{lv}$ is calculated experimentally at $T_c$, and values for $\gamma_{sl}$ are calculated at $T_c$ by determining interfacial entropy via radial distribution functions ($\Delta \gamma = \gamma_{sv} - \gamma_{lv} - \gamma_{sl}$). The value of $\xi$ is calculated from equation \textbf{A.14} in \cite{pluis1990surface} for the value of $T_c$ from the given potentials and $\gamma_{sl}$ above.}
\end{flushleft}
\end{table}

The equations of motion were integrated using a Verlet method, with an integration timestep of $2.0$ fs for nickel and $1.5$ fs for aluminium. The interaction between nickel atoms were modelled using an embedded-atom-potential (EAM) developed by Sheng et al \cite{sheng2011highly}. For aluminium we used the Ercolessi et al \cite{liu2004aluminium} glue potential. These potentials give melting temperatures of $1650$K and $930$K for Ni and Al respectively. The melting temperature for the Ni potential is lower than the real melting temperature of nickel (1728 K), but this is typical for EAM potentials. The Al potential gives a better estimate of the real melting temperature (933 K). This potential has been used previously to study surface melting in Al nanoparticles \cite{hendy2009superheating}.

As noted, EAM potentials typically underestimate the bulk melting temperature, but simulated values of $\gamma_{lv}$ and $\gamma_{sl}$ may also differ from experimental values. Nonetheless, determining these values experimentally, or even via simulation, is difficult. Indeed, values of these quantities reported in the literature vary appreciably, so the values reported in table~\ref{tab:critical_R} should be taken as indicative only. For our purposes, we wish to observe wires with melting and non-melting surfaces, and in subsequent sections we will illustrate that this is the case with these potentials.



\subsection{Identifying solid and liquid atoms}
\label{SS:Comp_q4q6}
The structure of the solid and liquid components of the wire were classified by using the local bond order parameter $q_6(i)$ introduced by Steinhardt et al \cite{steinhardt1983bond} and then taking its modified, average value $\bar{q}_6(i)$ \cite{lechner2008accurate}. This parameter is sensitive to different crystal structures and is a measure of the local symmetry around an atom. With this local bond order parameter it becomes possible to resolve if an atom is in a more `solid-like' or `liquid-like' state. The distribution of $\bar{q}_6(i)$ becomes bimodal when looking at a coexisting solid-liquid state, with one peak appearing at $\bar{q}_{6} \approx 0.491$ for `solid' FCC atoms and another peak appearing at $\bar{q}_{6} \approx 0.181$ for `liquid' atoms. From this we can take the average value of the peaks in the distribution and define a cutoff value, $\bar{q}_{cut}$ which for FCC lattice structures is $\bar{q}_{cut} \approx 0.33$. So if $\bar{q}_{cut} > 0.33$ the atom is more likely to be solid and if $\bar{q}_{cut} < 0.33$ then the atom is more likely to be liquid. With this a `state' can be assigned to each atom based upon the order parameter $\bar{q}_{6}$. Note that the $\bar{q}_6$ parameter exaggerates the number of `solid' atoms due to sensitivity to the local order in the liquid state. Atoms in the liquid state and atoms around the interface fluctuate between solid and liquid more frequently, so computing the average-state of each atom over a small time interval more precisely classifies each atom according to its phase it is surrounded by. This eliminates isolated atoms identified being instantaneously identified as solid in the interior of the liquid melt, and more clearly defines the solid at the solid-liquid interface.

\subsection{Surface melting and melting temperature simulations}
\label{SS:Comp_Melting}
Caloric curves were constructed to obtain melting temperatures for each metal for a range of wire radii. For nickel, wires had radii ranging from $R = 7$\AA\ to $R = 90$\AA\ and a constant length of $H=141$\AA. For aluminium wires, the radius ranged from $R=8$\AA\ to $R=100$\AA\ with a constant length of $H=162$\AA. A Langevin thermostat \cite{schneider1978molecular} was used to control the temperature with damping parameter of $0.1$ps$^{-1}$. For melting simulations of nickel, runs took place over $4.0$ns with a heating rate of $200$K/ns. For aluminium, runs for smaller wires (\textit{i.e. $R<32$\AA}) took place over $10.0$ns with heating rates around $55$K/ns, and for larger wires ($R>32$\AA) were run over $4.0$ns with a heating rate of $120$K/ns.
 The melting temperature $T_m$ is the temperature where the most prominent peak in the specific heat capacity at constant volume $C_v = \frac{dE}{dT}$ as the wire is slowly heated by the thermostat. In contrast, the surface melting temperature $T_s$ was found for each facet ($\{100\}$ or $\{110\}$) by identifying the temperature where the mean value of $\phi_L$ on this facet exceeded $0.5$.

\subsection{Wire breakup simulations}
\label{SS:Comp_Wire-breakup}
To study the breakup of the nanowires, wires of radii $21.1$, $28.2$, and $42.2$\AA\ with a length of $213$\AA\ were chosen for nickel, containing 31029, 54985, 123225 atoms respectively. Radii for aluminium wires were $24.3$, $32.4$, $48.6$\AA\ with a length of $243$\AA\ with each containing 32737, 57181, 126397 atoms respectively. These were chosen to give a range of different aspect ratios for each wire and to take note on how the evolution of the melting dynamics changed as the aspect ratio changed.
Individual runs were carried out to study the melting dynamics and wire breakup for each wire radii. We again used a Langevin thermostat to control the temperature with a larger damping parameter, $10$ps$^{-1}$, to slow the break-up process. Physically this would correspond to a wire in weaker thermal contact with its environment. Ni wires were heated the wires from around 1350 to 1550K over a period of 4.0 ns (i.e. a heating rate of $50$K/ns), while Al wires were heated from 700 to 1000K over a period of 4.0ns (i.e. a heating rate of $75$K/ns).

\section{Simulated Results}
\label{S:Sim_Res}
\subsection{Nickel}
\subsubsection{Surface melting and melting temperatures}
\label{Sec:Tm_and_Ts_Ni}
\begin{figure}[htp]
\resizebox{\columnwidth}{!}{\includegraphics{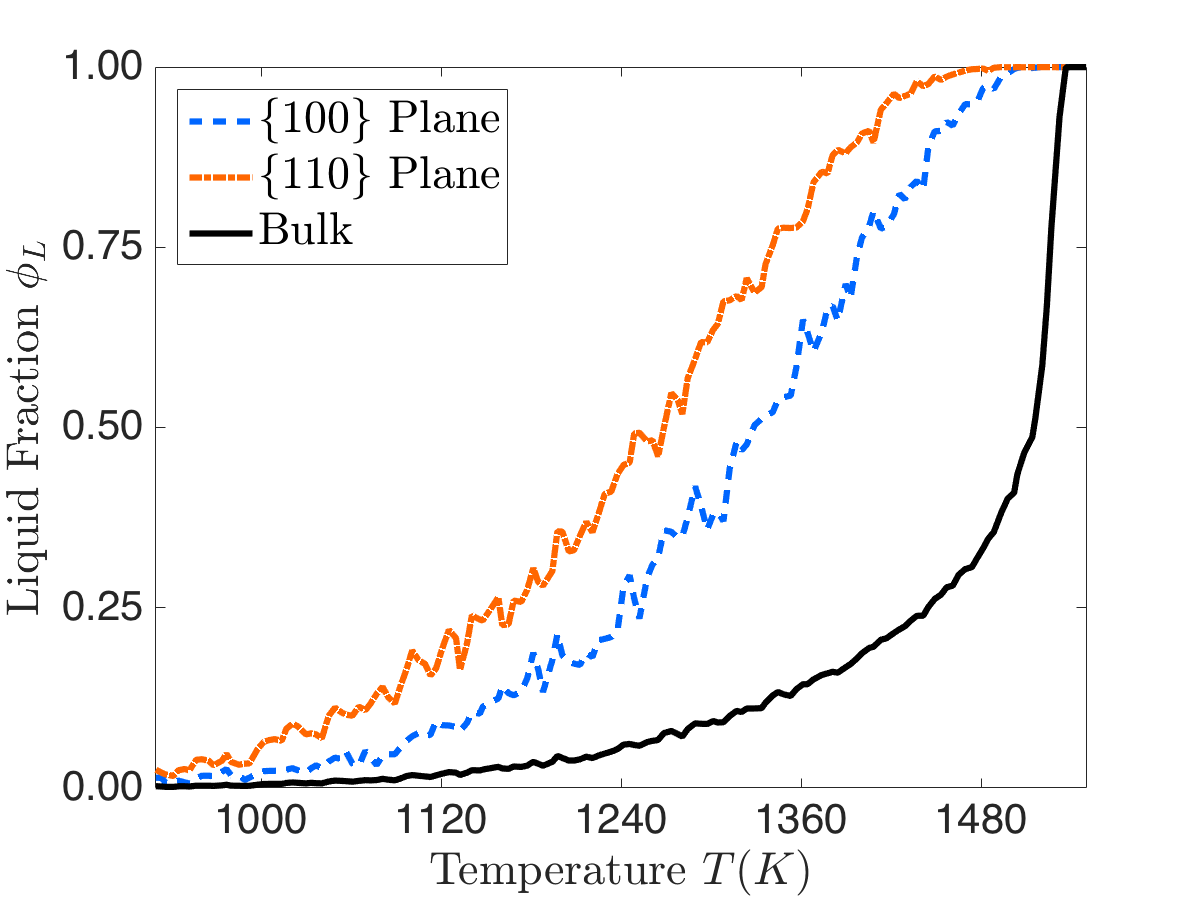}}
\caption{\label{Fig:Ni_R8_Liq_Frac} The liquid fraction $\phi_L$ is plotted against temperature for a nickel nanowire of radius $R=28.2$\AA\ and length $H=141$\AA\, for the bulk plus the $\{100\}$ and $\{110\}$ planes, illustrating the onset of surface melting and bulk melting.}
\end{figure}

We begin by calculating the melting point $T_m$ and surface melting point $T_s$ for nickel wires with a range of diameters using molecular dynamics simulations as described above. For the EAM potential used here, it is known that $\gamma_{sv} \{110\} > \gamma_{sv} \{100\}$ \cite{sheng2011highly}, so we would tend to expect that $T_s \{100\} > T_s \{110\}$.
Figure~\ref{Fig:Ni_R8_Liq_Frac} shows the surface and bulk liquid fractions $\phi_L$ for a wire with radius $R=28.2$\AA\ as a function of temperature. From the figure, it can be seen that surface melting initiates on the $\{110\}$ planes, and is followed by melting of the $\{100\}$ planes, before complete melting of the wire occurs. Thus we find that $T_s \{100\} > T_s \{110\}$ as suggested by (\ref{eqn:Tsurf}). Furthermore, the fact that surface melting initiates on both planes prior to complete melting suggests that $\Delta \gamma > 0$ for each plane.

\begin{figure}[htp]
\resizebox{\columnwidth}{!}{\includegraphics{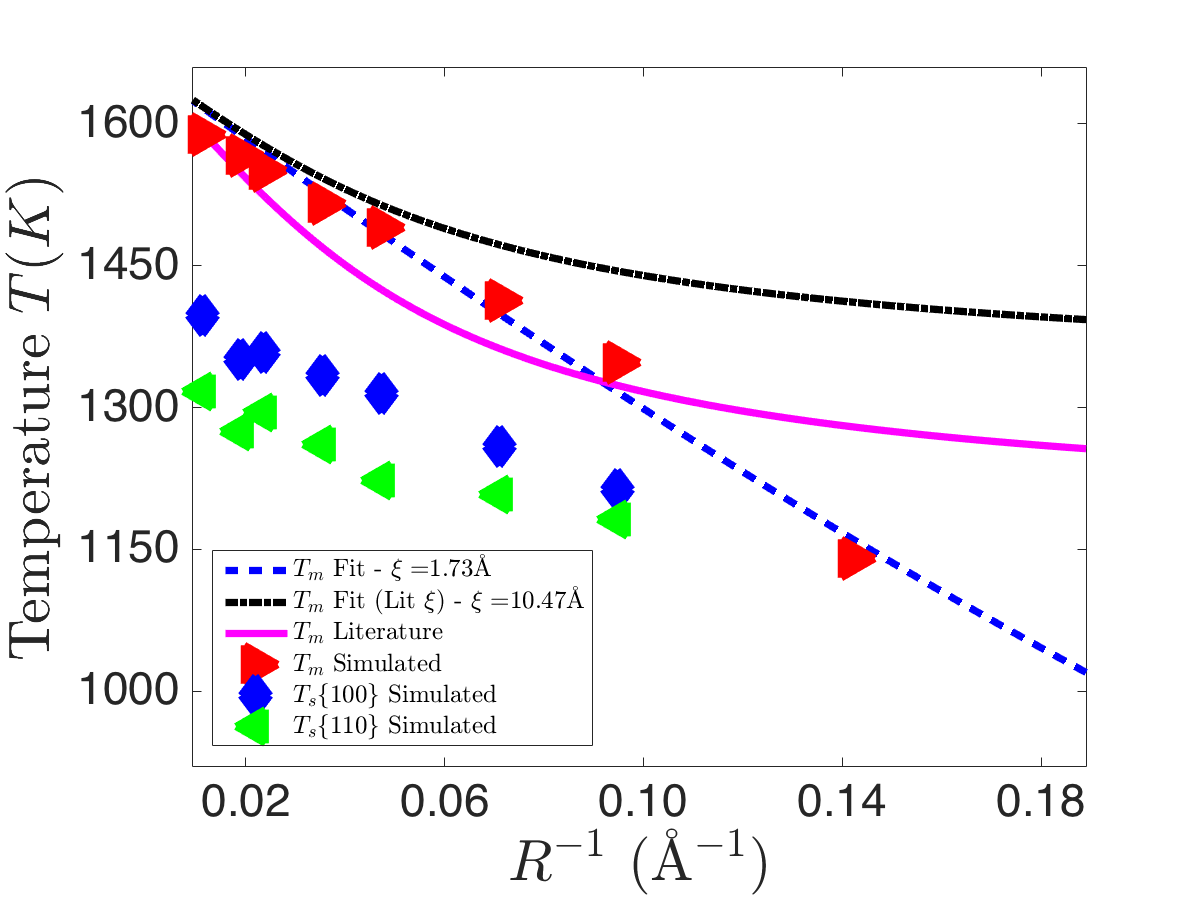}}
\caption{\label{Fig:Tm_Nickel} The simulated melting point temperatures $T_m$ and surface melting points $T_s$ for the $\{100\}$ and $\{110\}$ surfaces of nickel plotted against the reciprocal radius $1/R$. Also shown are the fits computed from equations \ref{Eq:fit1} (dashed line) and the fit when using the theoretical correlation length (dot-dashed line) as well as the equation~\ref{Eq:fit1} with literature values (solid-line).}	   
\end{figure}

Figure~\ref{Fig:Tm_Nickel} shows the values of $T_m$ estimated from the simulations for a series of wires of radius from 7 \AA \ up to 90 \AA \, as well as fits to these estimates using \cref{eqn:Tmelt}. The fitted curves for $T_m$ are of the form
\begin{equation}
T_{m}^{fit} = T_c \left(1 - \frac{4\lambda}{\rho LR}\left( \frac{I_0(R/\eta)}{I_0(R/\eta) + \phi I_1(R/\eta)}\right)\right)
\label{Eq:fit1}
\end{equation}
where for the potential used here, $T_c = 1650$K \cite{sheng2011highly} and $\lambda$, $\eta$, and $\phi$ are treated as fitting parameters. Comparing equation~\ref{Eq:fit1} and ~\ref{eqn:Tmelt} we can see that $\lambda = \gamma_{sl}$, $\eta = \xi$ and $\phi = \kappa = \frac{1 - \Delta \gamma/\gamma_{sl}}{1 + \Delta \gamma/\gamma_{sl}}$. From $\lambda$ and $\phi$ (\textit{i.e.} $\gamma_{sl}$ and $\kappa$) we can also calculate $\Delta \gamma$. Appropriate bounds were placed on each of the fitting parameters which would return estimates that were physically reasonable. Table~\ref{tab:Ni_Fit_Data} provides the values of the fitted parameters $\lambda$, $\phi$, and $\eta$, alongside a comparison to the parameters calculated using values obtained from the literature. 

The fitted values of $\lambda$ and $\phi$ are similar to those using values from the experimental literature, but the resulting values of $\Delta \gamma$ and $\xi$ are quite different.  Note that the values obtained from the fits for $\Delta \gamma$ are consistent with the observation that $\Delta \gamma >0$ for the wire surface facets. The atomic-scale ($\sim$ 2 \AA) value of $\xi$ from the fits to $T_m$ is interesting. While $\xi$ is an effective parameter that describes the average thickness of the solid-liquid interface, in (\ref{eqn:Tmelt}) it causes $\xi/R$ deviations from the classic $1/R$ behaviour of melting temperature. The fact that the simulated melting temperatures do show a very nearly $1/R$ dependence on wire radius drives the best fits towards small values of $\xi$. Whether these small values of $\xi$ are physical or reflect a deficiency of (\ref{FEeqn}) is discussed further below.

\begin{table}[htp]
\begin{tabular*}{\columnwidth}{@{\extracolsep{\stretch{1}}}*{7}{r}@{}}
\hline
    		Quantity	\vline
	&\hspace{0.2cm} $\lambda$ 
    &$\phi$ \hspace{0.2cm}			&$\Delta\gamma$
    &$\eta$( \AA\ ) \\
\hline
Bulk \vline
&0.0145	
&0.782	
&1.50$\times10^{-3}$		
&1.73 \\
Literature	\vline		 
&\cite{gammas_sl_exp}0.0217
&\cite{tyson1977surface,gammas_sl_exp,gammas_lv} 0.570
&\cite{tyson1977surface,gammas_sl_exp,gammas_lv}5.90$\times10^{-3}$
&\cite{pluis1990surface}10.5
     \\
\hline
\end{tabular*}  
\captionsetup{justification=raggedright}
\caption{\label{tab:Ni_Fit_Data}
The table contains the values of the quantities extracted from equation~\ref{Eq:fit1}
with values calculated from the literature. The values for $\Delta\gamma$ were extracted from the values of $\phi$ numerically. Literature values were calculated using the values stated in table~\ref{tab:critical_R} from the aforementioned definitions of $\lambda$, $\phi$ and $\eta$ (the fitted value of $\xi$).}
\end{table}

The phenomenological model also suggests that there is a critical wire radius $R_c$ (given by equation~\ref{Eq:Rc_Fits}) below which bulk melting precedes surface melting. In figure~\ref{Fig:Tm_Nickel}, the surface melting temperatures for each facet and the bulk melting temperature converge at $1/R \approx 0.14$\,\AA$^{-1}$, suggesting that a critical radius $R_c$ lies between $R \approx 7-12$\AA. 

While equation~\ref{Eq:fit1} fits the MD data for $T_m$ well, albeit with unexpectedly low values of $\xi$, attempts to fit to equation~\ref{eqn:Tsurf} to the simulated $T_s$ produced poor results. Although the model described in the previous section provides an adequate description of the size dependence of the bulk melting temperature, it does not quantitatively describe the facet dependent surface melting temperatures for the considered wires. This is not surprising, given that the phenomenological model assumes an isotropic value $\Delta \gamma$ over the wire surface. However, the order in which the different surface facets melt is as would be expected from \Cref{eqn:Tsurf} if one were to treat each facet independently using this equation.

\subsubsection{Wire breakup}\label{Sec:Breakup_Ni}
\begin{figure}[htp]
\resizebox{\columnwidth}{!}{\includegraphics[width=0.30\textwidth]{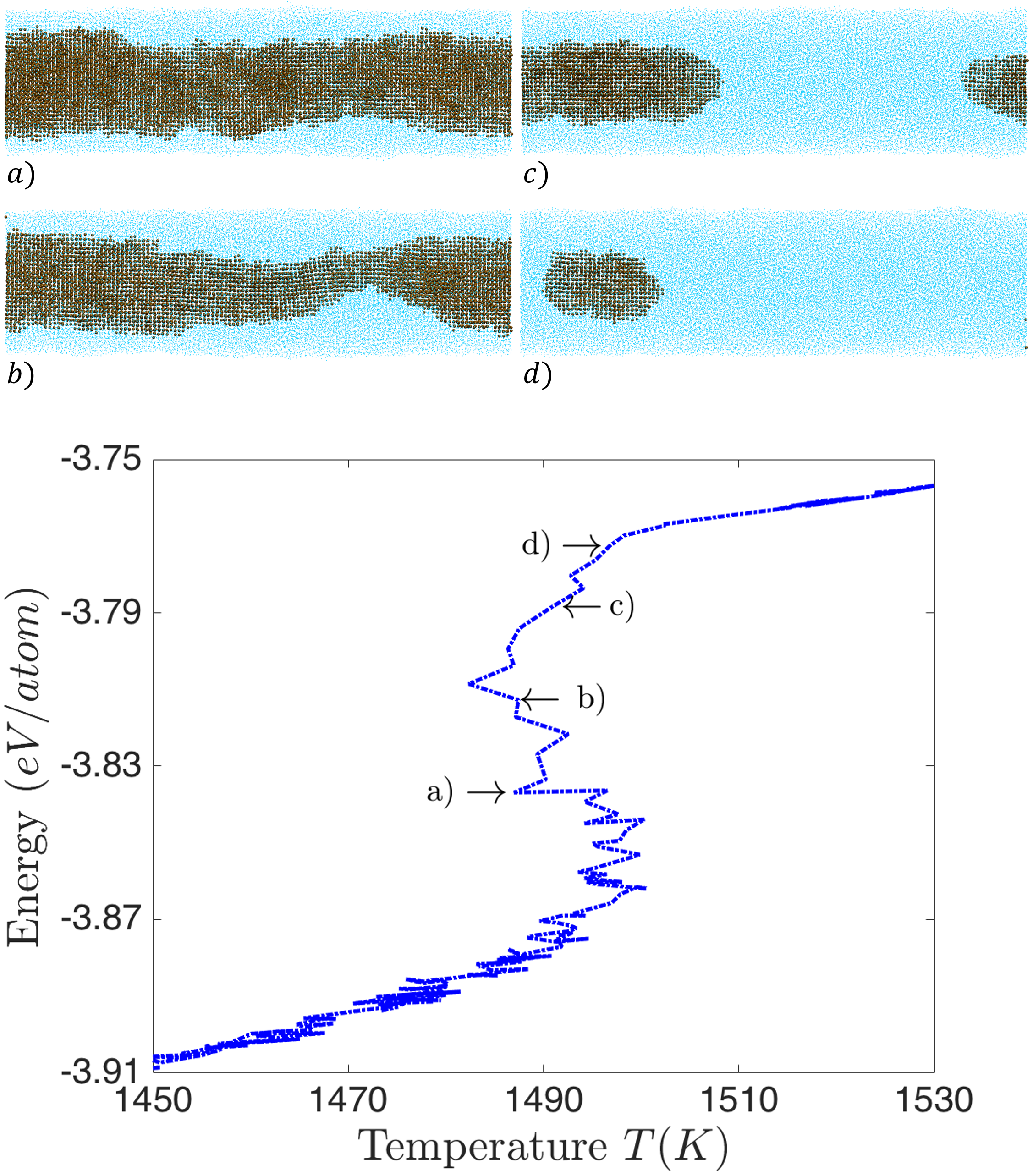}}
\caption{\label{Fig:Ni_Cal_Snaps} Snapshots of the $R = 28.2$\AA\ and $H = 213$\AA\ nickel nanowire looking down the $\{100\}$ plane, alongside its caloric curve. The wire undergoes complete melting at $T\sim$1500 K, Solid atoms are coloured brown (darker), and liquid atoms are coloured blue (lighter).}
\end{figure}

We now wish to investigate the effects of surface melting on wire breakup, as is shown in \Cref{Fig:Ni_Cal_Snaps}. This figure shows the caloric curve for a $R = 28.2$\AA\ nickel wire as it is heated to melting at $T\sim$1490 K, along with several snapshots from the simulation taken during the melting process. By the the time the bulk melting temperature is reached surface melting is fully developed, as would be expected from our considerations above, with the solid core completely covered with a thin layer of melt (\Cref{Fig:Ni_Cal_Snaps} a). However, prior to complete melting, the solid core pinches off and breaks (\Cref{Fig:Ni_Cal_Snaps} b), before the solid relaxes to form a spherical remnant (\Cref{Fig:Ni_Cal_Snaps} c and d) that finally melts. 

\begin{figure}[htp]
{\includegraphics[width=0.30\textwidth]{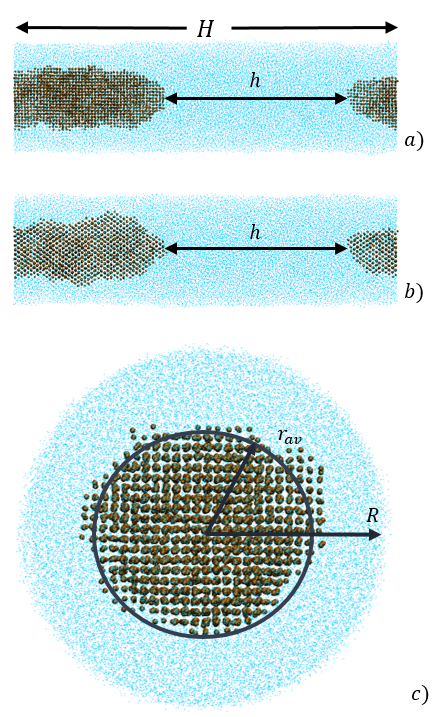}}
\caption{\label{Fig:Ni_rh_def}Snapshots of a nickel nanowire (the same seen in figure~\ref{Fig:Ni_Cal_Snaps}) with a radius $R=28.2$ \AA\ and length $H=213$ \AA during the complete melting process. The solid atoms are coloured brown (darker) and the liquid atoms are coloured blue (lighter). The top snapshot $a)$ looks along the $\{100\}$ direction of the solid wire, $b)$ is looks down the $\{110\}$ direction and $c)$ being a length-wise view in the $\{001\}$ direction. The radius of the solid $r$ is calculated by averaging along the length of the wire.. The liquid gap $h$ is calculated by taking the distance between the ends of the solid fragments.}
\end{figure}

To characterise the breakup process, we define a liquid gap width $h$ according to \Cref{Fig:Ni_rh_def}. Similarly we define an average solid core radius $r_{av}$ as shown in the figure. This is found by taking the solid atoms at the surface throughout the length of the wire, and then at each segment of the wire along its axis, taking an average value of the radius for each surface atom, then averaging along the wire length. On average the corresponding cylinder contains the outer surface of the solid core along the length of the wire. Of course, the actual radius $r$ varies along the length of the wire and it is evident that the solid core surface is anisotropic. Nonetheless, for our purposes $r_{av}$ gives a reasonable estimate of the solid core dimensions.

\begin{figure}[htp]
\resizebox{\columnwidth}{!}{\includegraphics{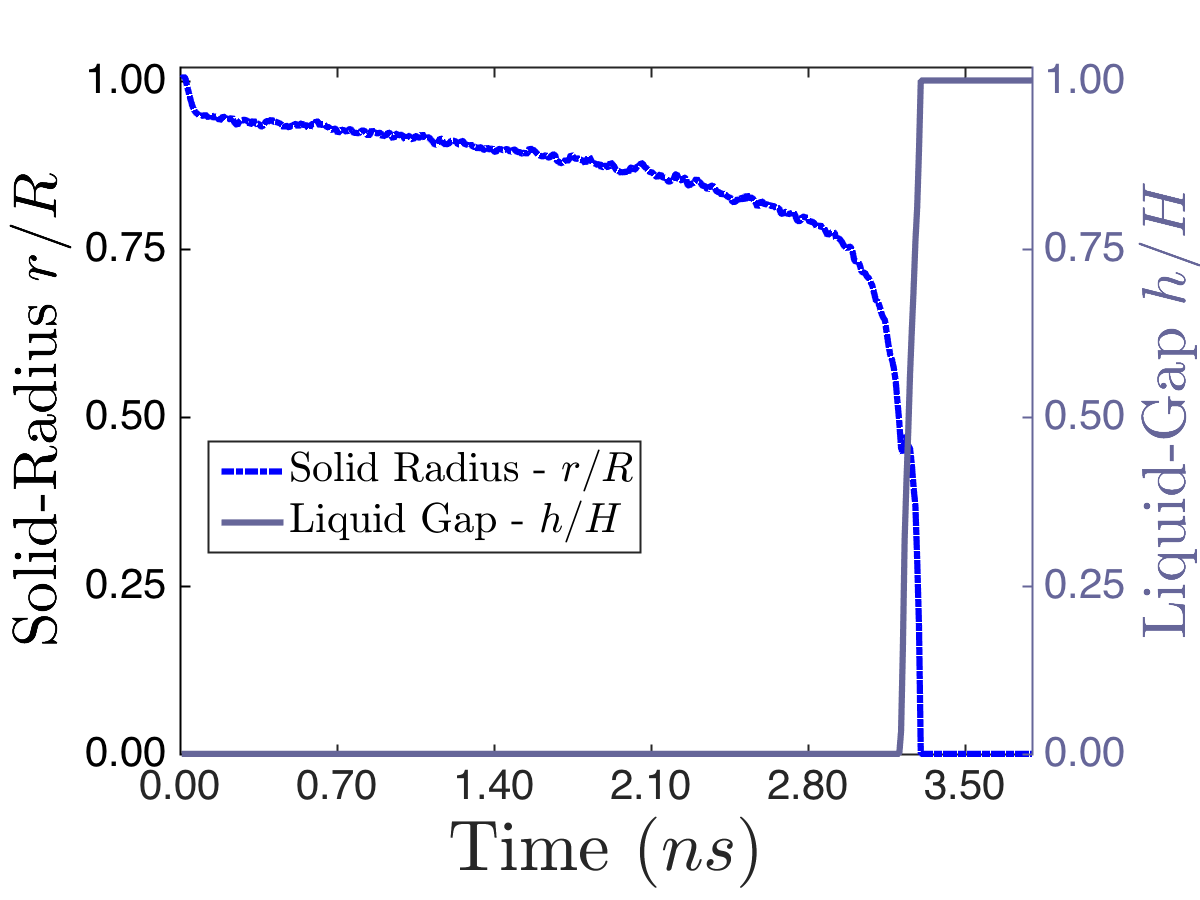}}
\caption{\label{Fig:Ni_R8_rh_g10_time} Time evolution of the solid radius $r$ and liquid gap $h$ for a nickel nanowire of radius $R=28.2$\AA\ and length $H=213$\,\AA. The initial temperature of the wire is $1350$K with a heating rate of $50$K/ns over $4.0$ns. The transition here takes place over about 3.8 ns, and the appearance of the liquid gap $h$ until complete liquefaction is over a period of about 0.096 ns.}
\end{figure}

\begin{figure}[htp]
\resizebox{\columnwidth}{!}{\includegraphics{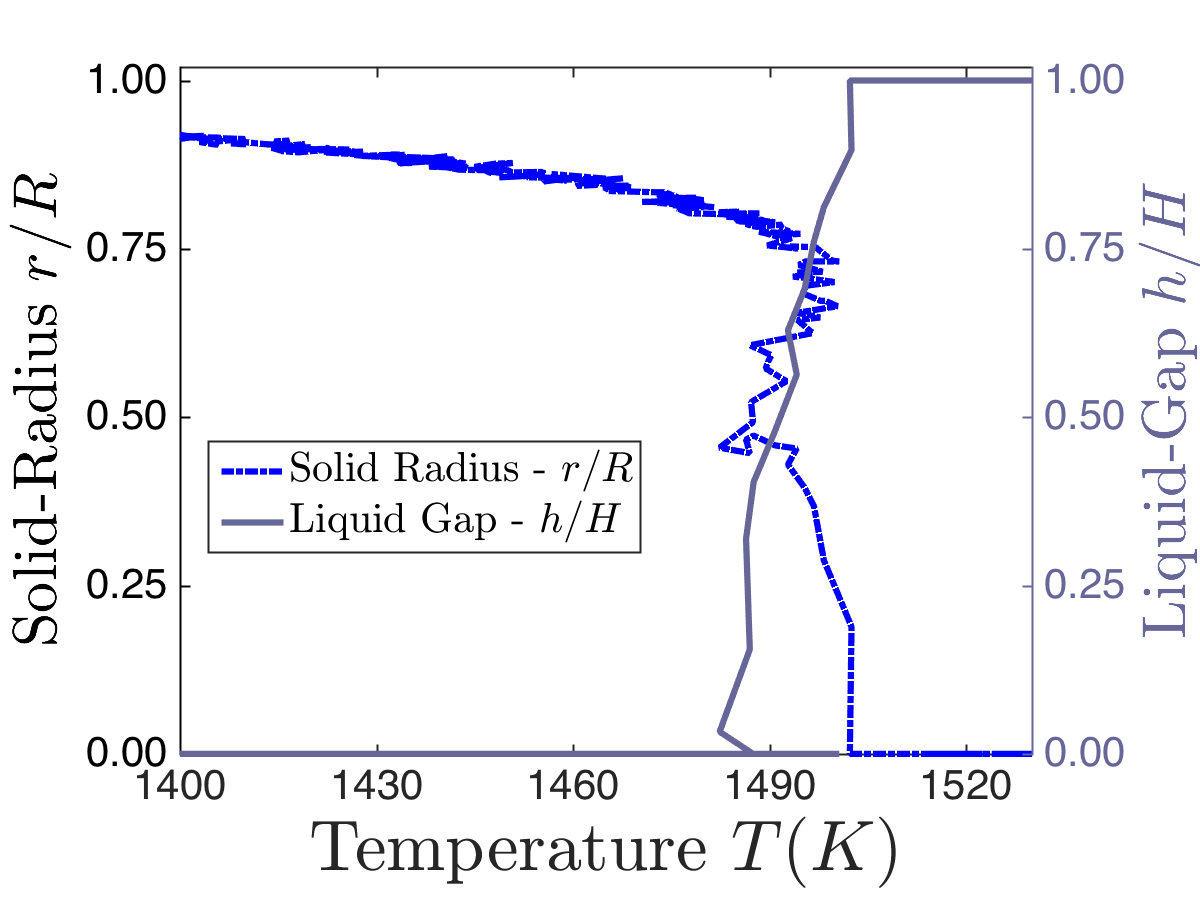}}
\caption{\label{Fig:Ni_R8_rh_g10_temp} The temperature evolution of the solid radius $r$ and liquid gap $h$ for a nickel nanowire of radius $R=28.2$\AA\ and length $H=213$\AA. The same initial temperature and heating rate mentioned in figure~\ref{Fig:Ni_R8_rh_g10_time} applies here. It can be seen that surface melting takes place here well prior to the solid being entirely consumed.}
\end{figure}

Figures \ref{Fig:Ni_R8_rh_g10_time} and \ref{Fig:Ni_R8_rh_g10_temp} illustrate the evolution of $h$ and $r$ as a function of time and temperature respectively for the $R = 28.2$\AA\ nickel wire. We note from Figure~\ref{Fig:Ni_R8_rh_g10_time} that while the radius of the solid core evolves relatively slowly, the growth of the liquid gap is rather rapid, consuming the solid core within a tenth of a nanosecond after it first appears. The solid core break up is also seen to occur at, or very close to, the melting temperature of the wire. Note that the break-up of the solid core is reminiscent of a Rayleigh-type instability (as discussed in \cite{wu2015self}), where a cylinder of radius $R$ is unstable to perturbations in the surface of the cylinder of wavelength $\lambda >  2 \pi R$. Our wires are periodic, with $H = 213$, so we would expect instabilities to drive break-up only for $R < 34$ \AA\, and the fastest unstable modes to be present on when $r < 24$ \AA\,. Indeed, Plateau-Rayleigh breakup of a completely molten wire (radius $R = 21.1$\,\AA) was observed after complete melting. In thicker wires, the period of the wire $H$ is $\leq 2 \pi R$, which suppresses unstable modes that would lead to Plateau-Rayleigh break up of the molten wire, but not necessarily of the thinner solid cores. Furthermore, we note that the onset of the break-up of the solid core that is illustrated in Figures~\ref{Fig:Ni_R8_rh_g10_temp} and~\ref{Fig:Ni_R8_rh_g10_time} occurs when its radius reaches $r \sim 13.8$\,\AA. 

In a thicker wire under the same initial temperature and heating rate, with $R=42.2$\AA\, we observed complete surface melting as seen before. The solid wire breakup however occurs after a larger portion of the surface has melted, namely at $r \sim 13.2$\,\AA. In the case of the thinner wire, $R=21.1$\AA\, under the same conditions, the solid breakup initiates sooner than the two former cases. This takes place at $r \sim 11.9$\,\AA. Thus we find that the solid cores remain stable down to smaller radii in thicker wires, but it is worth remembering that our wires are not in equilibrium when break-up occurs. 

\subsection{Aluminium}
\subsubsection{Surface melting and melting temperatures}\label{Sec:Tm_and_Ts_Al}
\begin{figure}[htp]
\resizebox{\columnwidth}{!}{\includegraphics{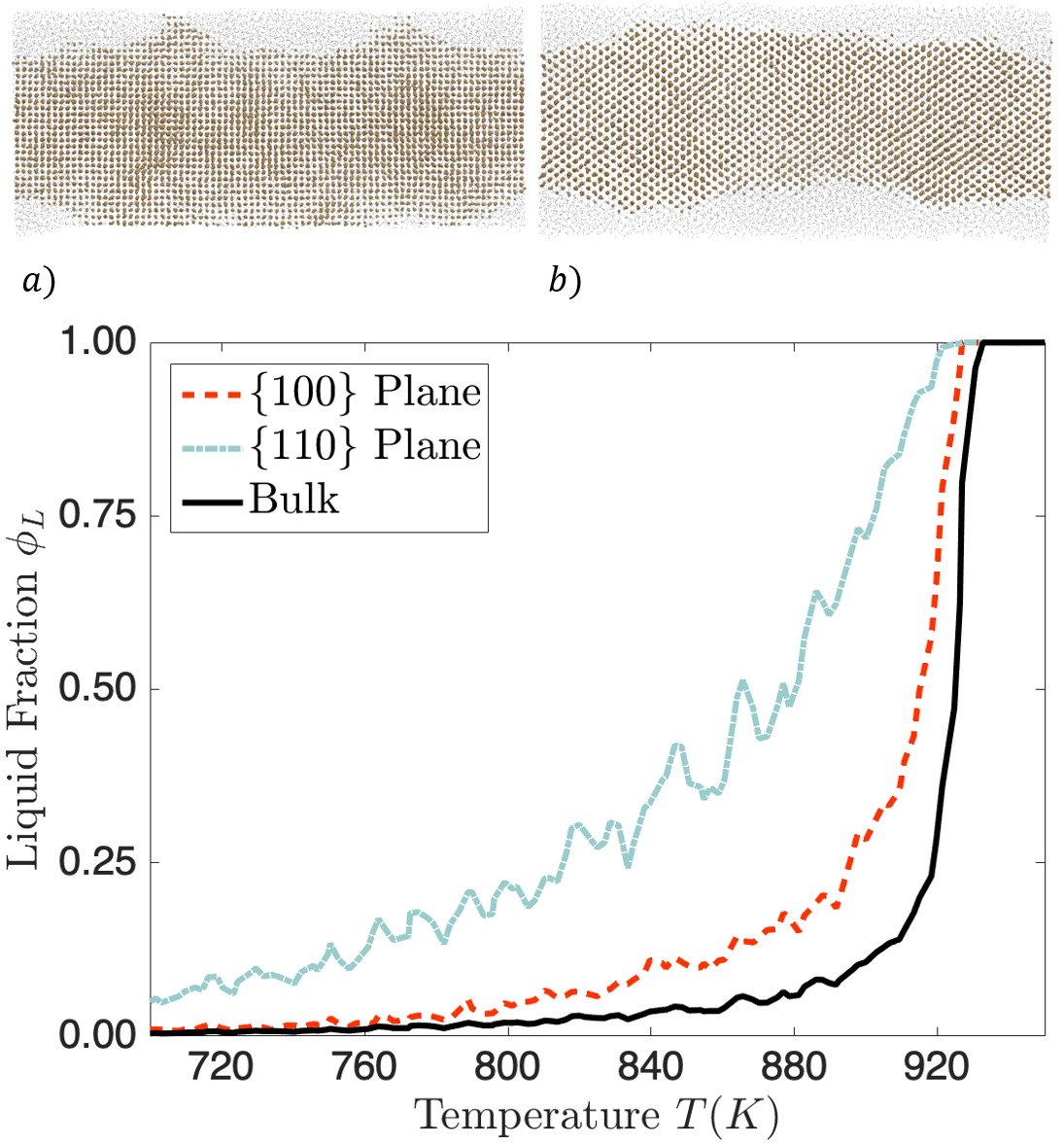}}
\caption{\label{Fig:Al_R8_Liq_Frac} The liquid fraction $\phi_L$ for the bulk and surfaces of an aluminium nanowire of radii $R=32.3$\AA\ and length $H=162$\AA\ as temperature is increased. Accompanying the liquid fraction plot are two snapshots viewing the \{100\} and \{110\} surfaces respectively. The darker atoms are solid and the lighter coloured atoms are Both snapshots are at the same instant in time at a temperature of $T \sim 920$K. The \{110\} facets melt well below $T_m$ while the \{100\} facets remain solid up to temperatures very close to $T_m$ and remain at least partially solid up to $T_m$.}
\end{figure}

We now consider the melting of aluminium wires with \{110\} and \{100\} surface facets. Experimentally, the Al\{110\} plane exhibits surface melting~\cite{van1990melting}, while the Al\{100\} plane has been observed to remain solid until the bulk melting temperature is reached~\cite{molenbroek1994anharmonicity}. We expect surface melting of an aluminium wire with melting and non-melting facets to play out differently to the nickel wire in the previous section. For the EAM potential used, we have $\gamma_{sv} \{110\} > \gamma_{sv} \{100\}$ \cite{liu2004aluminium}, so we would again expect from equation~\ref{eqn:Tsurf} that $T_s \{100\} > T_s \{110\}$ for aluminium. Figure~\ref{Fig:Al_R8_Liq_Frac} shows the surface and bulk liquid fractions $\phi_L$ for a wire of radius and length $R=32.4$\AA\ and $H=162$\AA\ as a function of temperature along with two snapshots of the wire viewing down the \{100\} and \{110\} axes respectively.


\begin{figure}[htp]
\resizebox{\columnwidth}{!}{\includegraphics{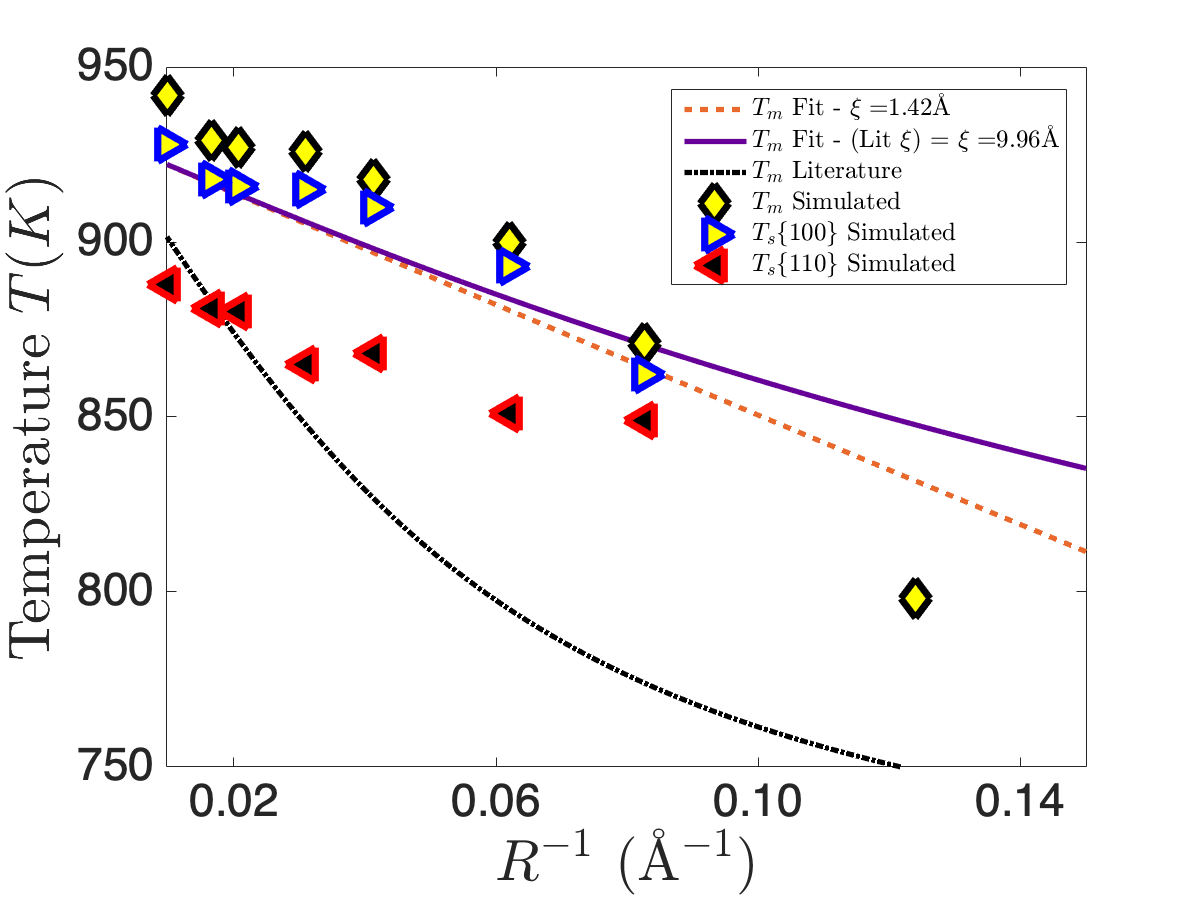}}
\caption{\label{Fig:Tm_Aluminium} The simulated melting point temperatures $T_m$ and surface melting points $T_s$ for the $\{100\}$ and $\{110\}$ surfaces of aluminium plotted against the reciprocal radius $1/R$. The fit for $T_m$ again computed from equation~\ref{Eq:fit1}.}
\end{figure}

These observations are consistent with $\Delta \gamma > 0$ for the $\{110\}$ surface, but the behaviour of the $\{100\}$ surface is more complicated. Past studies have shown that ordering persists on the Al$\{100\}$ surface up to its bulk melting temperature \cite{molenbroek1994anharmonicity}. Here we observe that while there are some atoms on the $\{100\}$ surface that are liquid, this liquid layer does not completely cover the surface until bulk melting commences, as seen in figure~\ref{Fig:Al_R8_Liq_Frac} and~\ref{Fig:Tm_Aluminium}. This suggests that $\Delta \gamma_{100}$ is close to zero.

As we did for nickel, we proceed by fitting Equation~\ref{Eq:fit1} to the simulated melting points of wires as a function of wire radius from $R\sim$ 10 \AA \ to 100 \AA. We fit to~\cref{Eq:fit1} and to~\cref{eqn:Tmelt} to estimate $T_c$, $\lambda$, $\eta$, and $\phi$ for aluminium. As can be seen in Figure~\ref{Fig:Tm_Aluminium}, the best fits are poor compared to the corresponding fits for the nickel. 
The fitted values of $\lambda$ and $\phi$ (\textit{i.e.} $\gamma_{sl}$ and $\kappa$) can be used to estimate $\Delta \gamma$, but these estimates are all an order of magnitude away from their corresponding literature values. Poor agreement is perhaps to be expected as the simulations reveal a highly anisotropic melting process, which is not reflected in the cylindrically symmetric model behind Equation~\ref{eqn:Tmelt}. 

Note that for this potential, the bulk melting temperature is 930 K, while for the largest wires $T_m$ exceeds this. This is very similar to the overheating observed in large aluminium clusters \cite{hendy2009superheating} that is associated with the the presence of non-melting surfaces. This may be why the change in melting temperature is not as well described by a deviation proportional to $1/R$, as is seen in Fig~\ref{Fig:Tm_Aluminium}. In the smallest wires, bulk melting is observed to occur without prior surface melting as was the case for nickel. In figure~\ref{Fig:Tm_Aluminium}, the surface melting temperatures for each facet and the bulk melting temperature converge at $1/R \approx 0.124$\,\AA$^{-1}$, suggesting that a critical radius $R_c$ lies between $8-12$ \AA.

\begin{table}[htp]
\begin{tabular*}{\columnwidth}{@{\extracolsep{\stretch{1}}}*{7}{r}@{}}
\hline
    		Quantity	\vline
	&\hspace{0.2cm} $\lambda$ 
    &$\phi$ \hspace{0.2cm}			&$\Delta\gamma$
    &$\eta$( \AA\ ) \\
\hline
Bulk \vline
&0.0101	
&6.92	
&-0.00750		
&1.42 \\		 
Literature	\vline		 
&\cite{gammas_sl_exp}0.00750
&\cite{tyson1977surface,gammas_sl_exp,gammas_lv} 1.12
&\cite{tyson1977surface,gammas_sl_exp,gammas_lv}-0.000630
&\cite{pluis1990surface}$9.96$
     \\
\hline
 	\end{tabular*}  
\begin{flushleft}
\captionsetup{justification=raggedright}
\caption{\label{tab:Al_Fit_Data}
A table containing the values for aluminium of the quantities obtained from fits to equation~\ref{Eq:fit1} with values calculated from the literature. The values for $\Delta\gamma$ were obtained from the values of $\phi$ numerically, while literature values were calculated by using the aforementioned definitions of $\lambda$, $\phi$ and $\eta$.}
\end{flushleft}
\end{table}

In summary we find that the isotropic model for the Al wires embodied in Equation~\ref{FEeqn} works even less well than for Ni wires. For the simulated Al wires, where we have observed both melting and non-melting surface facets, it seems that correctly modelling the resulting anisotropy must be a prerequisite for a quantitative description of wire melting.  

\subsubsection{Wire breakup}
\label{Sec:Breakup_Al}
\begin{figure}[htp]
\resizebox{\columnwidth}{!}{\includegraphics[width=0.30\textwidth]{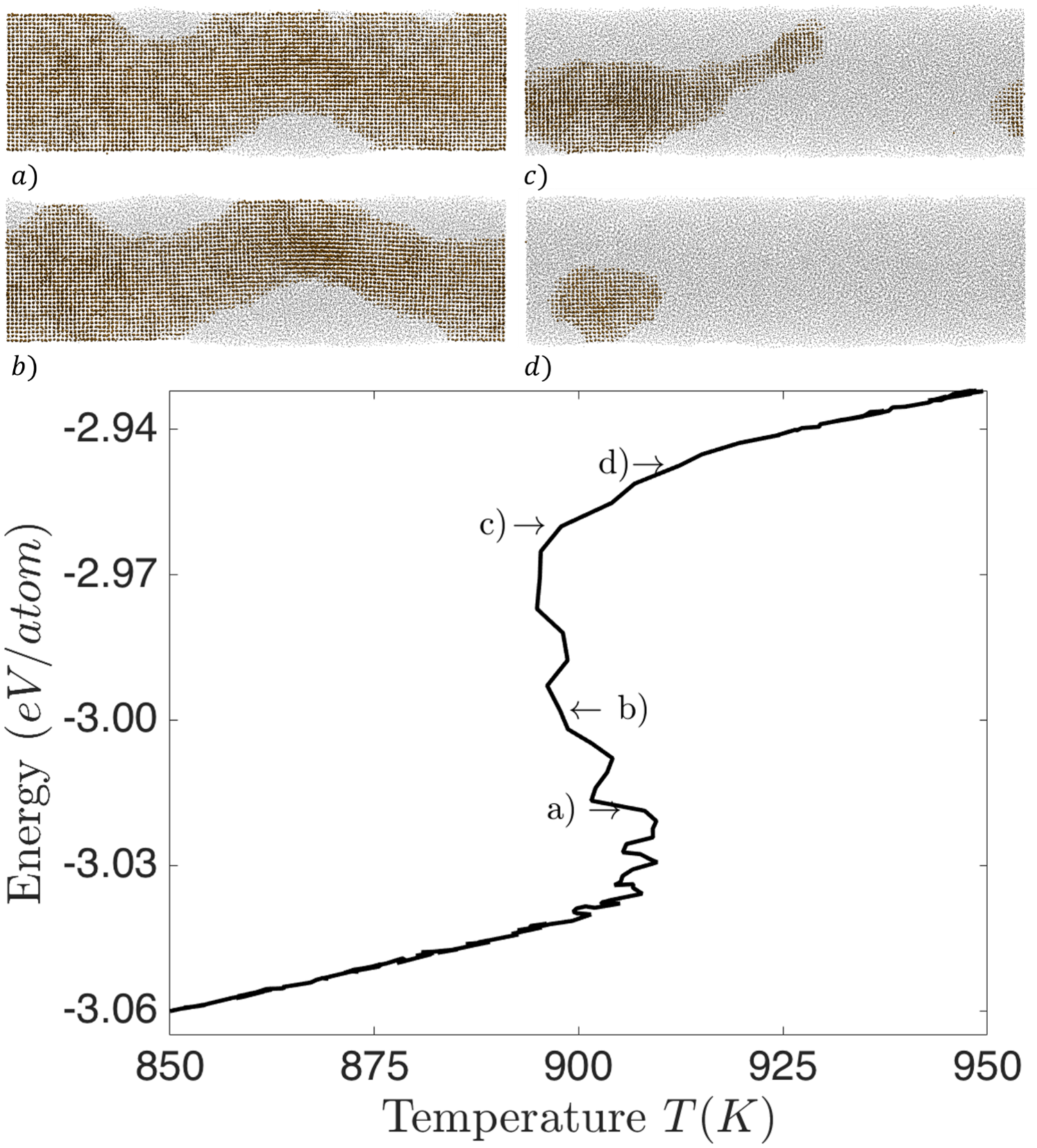}}
 \caption{Snapshots of the $R=32.3$\AA\, $H=243$\AA\ nanowire during the melting transition are shown, looking down the $\{100\}$ plane (highlighting the non-melting features), with the corresponding caloric curve below. Solid atoms are coloured gold (darker), and liquid atoms are coloured grey (lighter). It should be noted that in a)-d) there is only partial melting as viewed down the \{100\} plane, while surface melting is complete over the \{110\} surfaces. This is particularly evident in d) where even though the remaining solid is very small, a dry patch of a \{100\} surface plane still persists.}
\label{Fig:Al_Cal_Snaps}   
\end{figure}

We again investigate how surface melting effects the aluminium wire breakup. Turning our attention to Figure~\ref{Fig:Al_Cal_Snaps}, we have a caloric curve and some snapshots for a $R=32.3$\AA\ aluminium wire until it melts at $T\sim$900 K. In the figure here we see again that as the bulk melting temperature is reached there are still dry surfaces in each snapshot along the caloric curve even after the breakup has occurred. This is in contrast to nickel where complete surface melting develops when the melting transition begins. Looking down the $\{100\}$ direction in the snapshots we see small liquid pockets form on the $\{100\}$ surface (Figure~\ref{Fig:Al_Cal_Snaps} $a$). These liquid pockets spread from the \{110\} plane by melting higher indexed facets spreading towards the \{100\} surface. These regions expand as higher index planes at the solid-liquid interface melt (Figure~\ref{Fig:Al_Cal_Snaps} $b$). We then see the solid core pinch-off and melt (Figure~\ref{Fig:Al_Cal_Snaps} $c)$ and $d$) leaving behind a solid remnant with a small dry surface patch.

\begin{figure}[htp]
	\centering
{\includegraphics[width=0.30\textwidth]{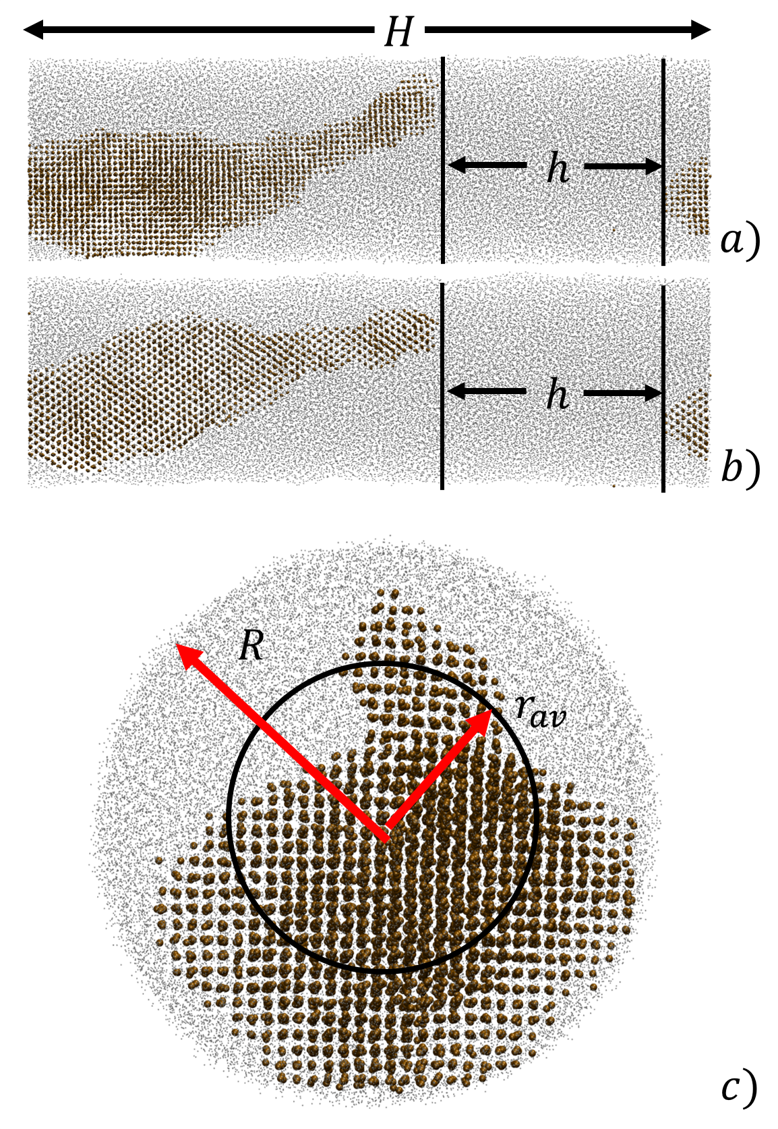}}
    \caption{An aluminium nanowire of radius $R=32.3$\AA\ and length $H=243$\AA\ at an instant in time. The solid atoms are coloured brown (darker) and the liquid atoms are coloured grey (lighter). Snapshots are taken of the $\{100\}$ $\{110\}$ and $\{001\}$ planes respectively. The same technique as in \Cref{Fig:Ni_rh_def} was used to calculate $r$ and $h$, except in this case the average radius may be slightly exaggerated due to non-melting of \{100\} surfaces.}
	\label{Fig:Al_MD_Def}   
\end{figure}

As we have seen already in Figure~\ref{Fig:Al_Cal_Snaps}, the solid aluminium wire breaks up quite differently to the corresponding nickel wire. In this case both non-melted and melted surfaces are present up until complete melting. We now look more closely at this in figure~\ref{Fig:Al_MD_Def}, which shows how $r$ and $h$ evolve as a function of temperature for an aluminium nanowire of radius $R=32.3$\AA\ and length $H=243$\AA. As before, $r_{av}$ is calculated by taking the solid atoms at the surface throughout the length of the wire, and at each segment of the wire along its axis, taking an average of the radius of each surface atom, and then averaging along the wire length. 

This is shown in figure~\ref{Fig:Al_MD_Def}. Here we see that there are non-melted and melted surfaces present in the insets a) and b) when viewed down the $\{100\}$ and $\{110\}$ planes respectively. Looking down the $\{001\}$ plane in inset c) we see that the $\{110\}$ planes have largely surface melted, while there still remains dry $\{100\}$ surface. Due to this non-melting portion of the surface, the solid radius $r$ is as defined in inset c) is not a particularly good fit to the shape of the wire, but does provide an estimate of the radial extent of the remaining wire. 

\begin{figure}[htp]
\resizebox{\columnwidth}{!}{\includegraphics{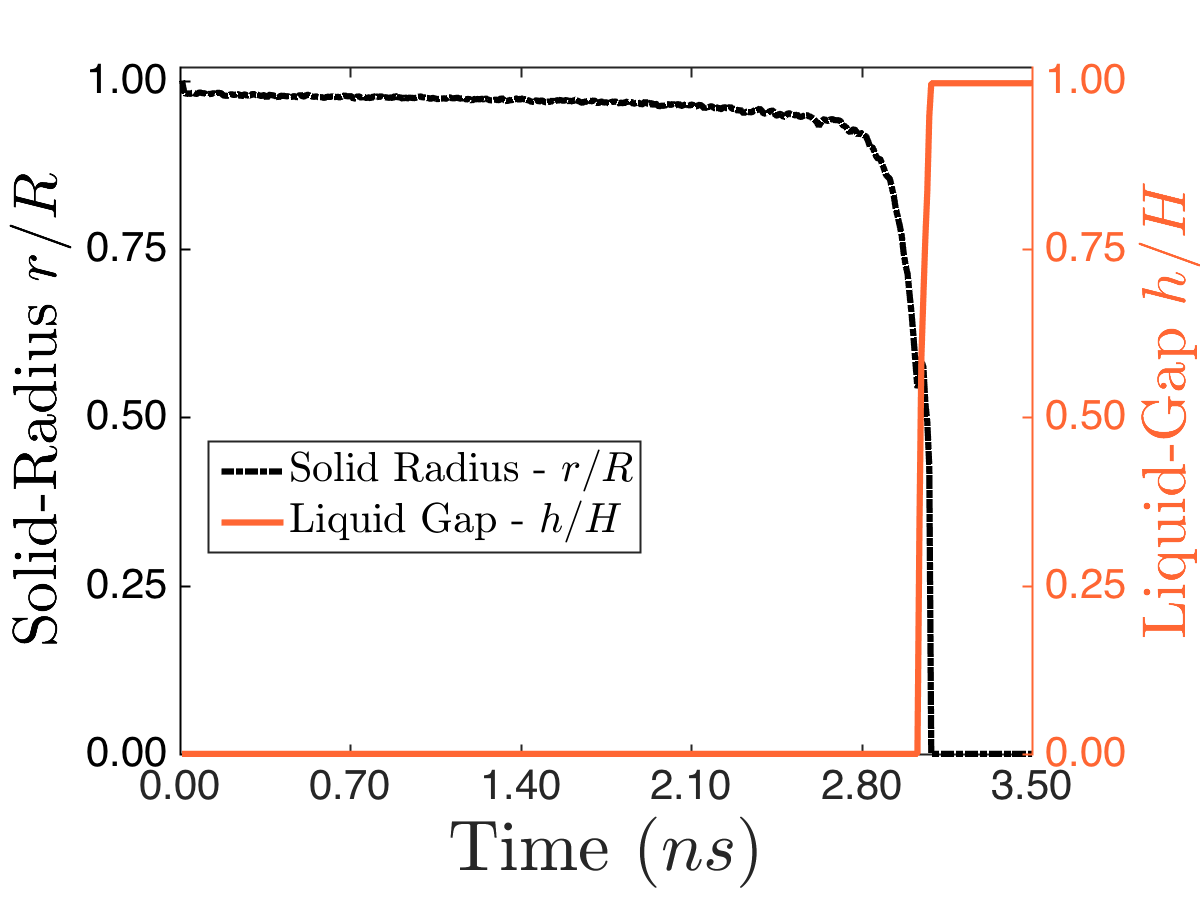}}
    \caption{Time evolution of the aluminium $R=32.3$\AA\, $H=242.0$\AA\ nanowire. The transition for aluminium seen here takes place over about $3.5$$ns$, and the appearance of the liquid gap $h$ until complete liquefaction is over a period of around $0.057$$ns$, considerably shorter than it was for Ni. The radius at which the solid radius $r$ drops away is around $r_c \approx 0.54R$, slightly higher than that of Ni.}
	\label{Fig:Al_R8_rh_g10_time}   
\end{figure}

\begin{figure}[htp]
\resizebox{\columnwidth}{!}{\includegraphics{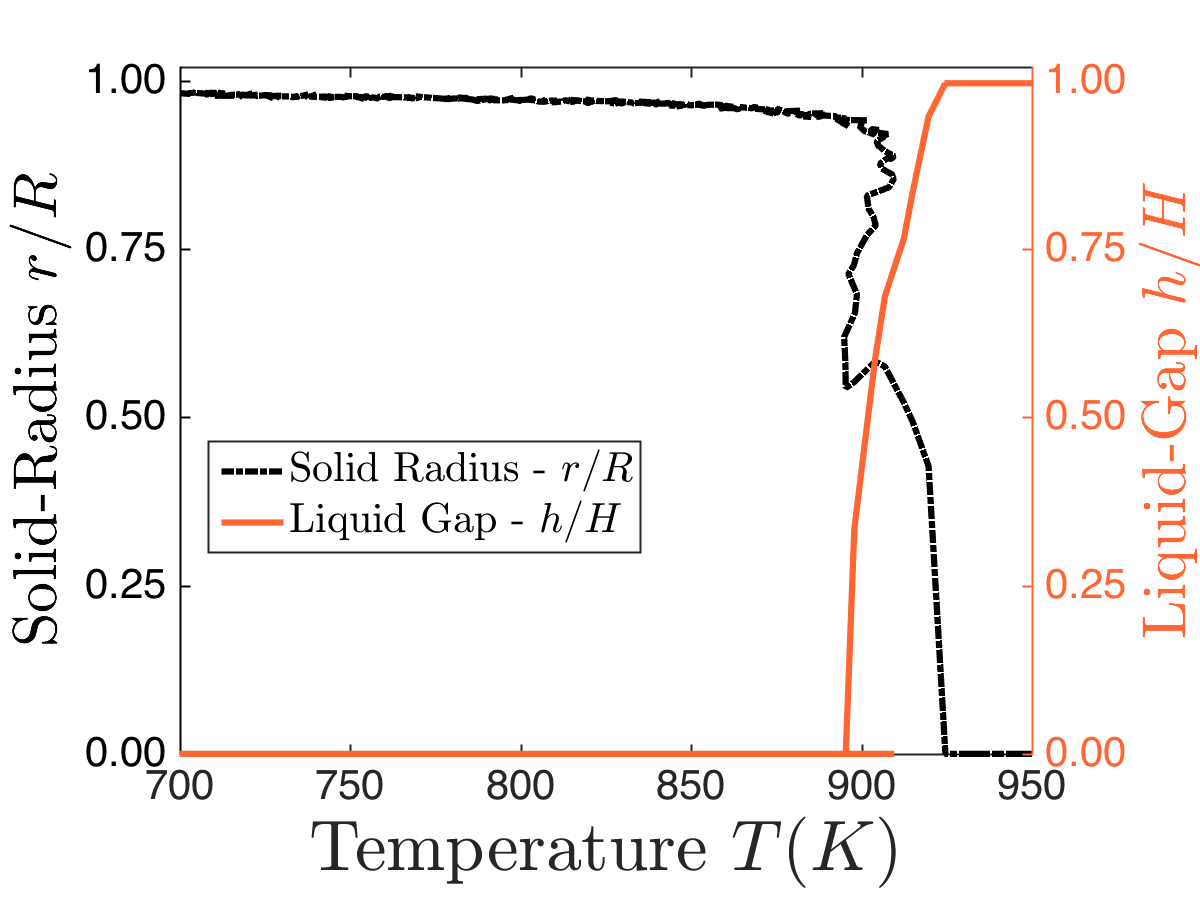}}
    \caption{Temperature evolution of the solid radius and liquid gap for the Al $R=32.3$\AA\, $H=243$\AA\ nanowire. The overheating effect pointed out earlier is especially obvious here again. It can also be seen here that significant surface melting doesn't take place until the onset of the melting transition, and the appearance of a breach in the entirety of the solid wire core.}
	\label{Fig:Al_R8_rh_g10_temp}   
\end{figure}

Figure~\ref{Fig:Al_R8_rh_g10_time} and \ref{Fig:Al_R8_rh_g10_temp} illustrate the how $r$ and $h$ for aluminium evolve as a function of time and temperature respectively. Comparing Figures~\ref{Fig:Ni_R8_rh_g10_time} and \ref{Fig:Al_R8_rh_g10_time}, we see the aluminium solid is consumed slightly faster than nickel. The radius of the solid core again occurs quite slowly with a rapid consumption of the solid once the liquid gap appears. The breakup of the solid core illustrated in figures~\ref{Fig:Al_R8_rh_g10_time} and~\ref{Fig:Al_R8_rh_g10_temp} occurs close to the melting temperature of the wire. The break-up of the solid core occurs once $r \sim 18$\AA, a slightly higher radius than that of the corresponding nickel wire. Aluminium wires here have a length of $H=243\AA$ where the fastest growing unstable modes would be present once $r < 27\AA$ according to Plateau-Rayleigh theory.

Examining thicker and thinner wires simulated under the same conditions, we observed that the thicker aluminium wires complete surface melting can occur prior to the onset of the breakup (e.g. with radii $R=48.6$\AA\ breakup occurs at $r \sim 22$\AA. For thinner wires the onset of the breakup occurs earlier than both the larger wires (e.g. with $R=24.3$\AA\ this takes place at $r \sim 18$\AA) while portions of the surface remain dry.

\section{Discussion}
\label{Sec:Dis}
Our simulations have shown that the two-parabola Landau model can describe bulk melting reasonably well for a nickel wire that has melting surfaces. The order in which the bulk and the surface facets melt is consistent with predictions of the model, despite the fact that it does not account for the anisotropy of the wire surface. However, to obtain quantitative agreement, it was necessary to invoke an atomically small correlation length ($\xi \sim 1 \AA$) that describes the width of the solid liquid interface. This is was not consistent with observations of the much rougher solid-liquid interface seen in many of the snapshots, but it is interesting that by including a more complex (albeit effective) description of that interface upsets the $1/R$ dependence of the melting temperature predicted by the theory. In one sense this is not surprising given that we normally understand that the $1/R$ dependence is a result of the changing surface area to volume ratio of the nanostructured material and this model complicates this picture. The surprise perhaps is that this dependence nonetheless persists in an anisotropic nanowire where interfaces are observed to be rough.  

For aluminium the Landau model performed poorly. For bulk melting qualitatively it produces the same sign of $\Delta \gamma$ as what is recovered from using values substituted from the literature and surface melting is described qualitatively by the model and the simulated results produce the correct order for melting. However, the stronger anisotropy present, with patches of non-melted surface persisting up until the complete melting of the wire, were associated with a more complex dependence of $T_m$ on wire radius. The cylindrically symmetric description of the free energy evidently does a poor job quantitatively of fitting the observed melting temperatures of the cylindrically asymmetric partially molten wires.  

For metal wires with melting surfaces (\textit{i.e.} if $\Delta \gamma > 0$), we would expect the model developed here to give an adequate quantitative description of melting and a qualitative description of surface melting, much as it did for Ni. For modelling materials bounded by non-melting facets (such as the Al(111) Al(100) and Pb(111) surfaces\cite{van1990melting,NM_surfs,pluis1987crystal}), however, the anisotropy should be taken into account. It may be that an anisotropic Landau model would be sufficient, but another approach is to incorporate elastic stresses at the solid-liquid interface \cite{levitas2011size,levitas2011coherent}. These models include features such as a non-zero shear modulus, which generate internal elastic stress at the interface, and a volumetric transformation strain, which allows for shape transformation during melting. Furthermore, to describe the break-up process it is likely that kinetic effects would also need to be accounted for. However these models can become quite complex and generally do not admit analytic solutions. The simpler model solved here it may be favourable to use to obtain some simple inferences of qualitative behaviour one may expect in experiments or simulation which has been adequately demonstrated in this study.

The anisotropy is also very visible in the breakup of the solid in the aluminium nanowires seen in section~\ref{Sec:Breakup_Al}. These wires break up via a very different pathway to the nickel wires. Note that the necking of the wire principally takes place along the \{110\} surface. Melting of the \{100\} plane takes place via liquid nuclei forming at arbitrary places along the surface which continue to grow by melting higher index crystal planes in an anisotropic manner. The thickness of the solid core at which breakup occurs is greater in nickel wires than in aluminium. This suggests a material dependence, possibly due to the fact that the wires are not in equilibrium, of the solid core breakup, which would not be expected in a Plateau-Rayleigh instability. This could be further investigated via a stability analysis of a time-dependent Landau-Ginzburg model, albeit with the caveats given above regarding the importance of anisotropy.

\section{Conclusion}
\label{Sec:Conclusion}

We have studied the bulk and surface melting of nickel and aluminium nanowires using molecular dynamics and a double-parabola Landau model. From the simulations, we observe the familiar $1/R$ dependence of these melting temperatures, something that could only be reproduced by the Landau model by assuming an atomic scale correlation length for the interface. In nickel, complete surface melting of the wire would precede the break up the sold core, which took place via the necking of the solid at a single point along the wire axis. In aluminium the pathway for solid core breakup was different, with complete surface melting occurring on the \{110\} plane and spreading in pockets of liquid to the \{100\} surface. Portions of the \{100\} surface remain solid up to the melting point of the entire wire. Our results illustrate the importance of surface anisotropy in the surface melting and melting of metal nanowires.

\bibliographystyle{apsrev4-1}
\bibliography{Melting}
\end{document}